\documentstyle[psfig]{l-aa} 
\begin{document}
\title{PEGASE: a UV to NIR spectral evolution model of galaxies}
\subtitle{Application to the calibration of bright galaxy counts}
\author{Michel Fioc \inst{1} \and Brigitte 
Rocca-Volmerange \inst{1,2}}
\thesaurus{3(11.05.2; 13.09.1; 09.04.1; 11.19.5; 12.03.2; 11.06.2)}
\offprints{M. Fioc, e-mail: fioc@iap.fr}
\institute{Institut d'Astrophysique de Paris, CNRS,
98 bis Bd. Arago, F-75014 Paris, France \and
Institut d'Astrophysique Spatiale, B\^at. 121, Universit\'e Paris XI, F-91405 
Orsay, France}
\date{Received ??; accepted ??}
\maketitle
\markboth{M. Fioc \& B. Rocca-Volmerange: PEGASE, galaxy spectral evolution}
{M. Fioc \& B. Rocca-Volmerange: PEGASE, galaxy spectral evolution}
\begin{abstract}
PEGASE ({\bf P}rojet d'{\bf E}tude des 
{\bf GA}laxies par {\bf S}ynth\`ese {\bf E}volutive, in French)
is a new spectrophotometric evolution model for starbursts
and evolved galaxies of the Hubble sequence.  
Its main originality is the extension to the near-infrared 
(NIR) of the atlas of synthetic spectra 
of Rocca-Volmerange \& Guiderdoni (1988) with 
a revised stellar library including cold star parameters
and stellar tracks extended to the thermally-pulsing regime
of the asymptotic giant branch (TP-AGB) and 
the post-AGB (PAGB) phase.
The NIR is coherently linked to the visible and the ultraviolet (UV),
so that the model is continuous on an exceptionally large
wavelength range from 220 \AA\ up to 5 $\mu$m.
Moreover, a precise algorithm allows to follow
very rapid evolutionary phases such as red supergiants or AGB crucial in the NIR.
The nebular component is also computed in the NIR. The extinction 
correction is gas-dependent for spirals and ellipticals.
A set of reference synthetic spectra at $z=0$, to which apply
cosmological k- and evolution e- corrections for high-redshift
galaxies, is built from fits of observational templates.
Because of the lack of visible to NIR spectral templates for each Hubble type, 
we adopt statistical samples of colors, not fitted by previous 
models. 

A first application of this continuous model 
is to solve the problem of the slope of the bright galaxy counts from $B=15$ 
to 19 and of the normalization parameter $\phi_{\ast}$ of the Schechter 
luminosity function.

Designed to be used currently and to take into account easily new
observational and theoretical inputs, the model is built as a series of independent
blocks. 
Any block is clearly identified and submitted to publication. Code sources,
input and output data are available\footnote{A first version is in the AAS 
CD-ROM Series, vol.~7 (Leitherer et al. 1996b)} by 
anonymous ftp at \mbox{\em ftp.iap.fr} in
\mbox{\em /pub/from\_users/pegase/} or at the WEB address of the authors:
\mbox{\em http://www.iap.fr/users/}.  
\keywords{Galaxies: evolution -- Infrared: galaxies -- dust, 
extinction -- Galaxies: stellar content -- Cosmology: 
miscellaneous -- Galaxies: fundamental parameters}
\end{abstract}
\section{Introduction}
The recent development of NIR observational astronomy,
as well as the progress
of the theory of stellar evolution (overshooting, nuclear rates, 
winds, opacities, see Maeder \& Meynet (1989); Chiosi (1986)), 
notably of the evolved phases (Groenewegen \& de Jong 1993; Vassiliadis \& 
Wood 1993) up to the PAGB terminal phases (Bl\"ocker 1995), put 
new constraints on the evolution of galaxies. 
The UV-optical range of currently star-forming galaxies 
is dominated by young stars, and results of models limited to this domain 
are unable to determine the past star formation history.
To break this degeneracy, the NIR emission of the bulk of giants
is worthy to be examined as a meaningful indicator of mass and age.
Moreover, the NIR light is less obscured by dust than at shorter wavelengths and
has proved very useful in the spectral analysis of dusty starbursts 
(Lan\c{c}on et Rocca-Volmerange 1996).
Following star formation history over a long timescale
thus needs a continuity of the wavelength range. That continuity is also 
necessary
to analyze distant galaxies at large redshift ranges in the deepest galaxy counts.
From the long series of models compared by Arimoto (1996) and Charlot (1996) 
at the Crete meeting, most current models agree in the visible. However, 
none of them is published in the NIR by Leitherer et al. (1996a),
possibly because they show a flux deficit from $J$ to $K$
in the spectral energy distribution (SED) of early-type galaxies 
(Arimoto 1996). The poor knowledge 
of the atmospheric parameters of cold stars dominating the 
NIR and the rapid evolution of the latest phases are 
evident difficulties for the modeling of NIR emission. 
Another difficulty is the connection 
of the NIR stellar emission  to the  visible at about 1 $\mu$m,
where cold  stars of $T_{\rm eff}\simeq$ 2000 to 3000 K peak. 
Moreover, in that domain, only a few data are available because 
receivers in the optical as in the NIR have minimal sensitivity.

Our main goal is to build a new atlas of evolving
standard synthetic spectra for the types of the Hubble sequence.
We present here UV to NIR energy distributions for 8 spectral galaxy types.
Evolutionary parameters are constrained by fitting synthetic spectra
at $z=0$ on integrated spectra and colors of nearby galaxies.
The choice of templates requires a peculiar attention 
to aperture effects and a significant identification of Hubble types.
These criteria have led us to use statistical color samples, 
because of the lack of significant spectral templates in the NIR 
wavelength range.
The samples of optical and NIR data show however a high dispersion,
which is partly observational, but is also due to uneven star formation 
histories leading to an intrinsic scatter of colors for a given morphological
type. This makes it difficult to identify observational templates.

A first application of PEGASE solves the puzzling question of the 
bright galaxy counts. The slope at bright magnitudes is shown to be 
in agreement with recent {\em multispectral} observations of Gardner et al. 
(1996), excluding the strong evolution of giant galaxies at low redshift 
advocated by Maddox et al. (1990) from their steep counts. Moreover, the 
normalization of the luminosity function is in accordance with the high value
of Marzke et al. (1994).

The structure of this paper is as follows. In section \ref{modele}, we
present our model, the algorithm, the stellar library and other related data, 
the evolutionary tracks, the nebular emission and finally the extinction
model.
In section~\ref{histoire}, we examine the star formation history
for starbursts and evolved galaxies and propose evolutionary scenarios fitted
on UV to NIR observations of galaxies of the Hubble sequence.
In section~\ref{comptages}, we check that the evolution of our 
standard scenario is realistic by comparison with bright galaxy 
counts, and we finally discuss in the conclusion the advantages and 
limitations of the model.
\section{The model}
\label{modele}
The two historical models of Tinsley (1972) and Searle et al. (1973) 
computed, respectively, the photometric evolution of galaxies
from isomass stellar evolutionary tracks and isochrones. When 
applied to the same stellar evolutionary model and with similar input 
data (stellar spectra library, bolometric corrections\ldots), both methods should 
give equivalent results, so that output differences must arise from deficient 
algorithms. Refined algorithms
recently improved to conserve the released energy and the stability of outputs 
without suffering any degradation by smoothing methods are presented here. 
The stellar library is
improved in the visible and the UV and is coherently
extended by using NIR spectra and colors of stars on a
significant cover of the HR diagram. Evolutionary tracks 
of the Geneva and Padova groups may be used up
to the beginning of thermal pulses and are completed with 
stellar models of the final phases up to the PAGB phase. 
The nebular emission, computed as in Guiderdoni \& Rocca-Volmerange (1987) 
(hereafter GRV), is extended to the
NIR and a new modeling of extinction in elliptical galaxies
is proposed. 
\subsection{The integration algorithm}
Although the principle of spectral evolutionary synthesis is simple,
computational problems and erroneous results may be caused by unoptimized
algorithms. 
The monochromatic flux of a galaxy at age $t$ and wavelength 
$\lambda$ may be written
\[F_{\lambda}(t)=\int_0^t\int_{m_{\rm l}}^{m_{\rm u}}\tau(t-\theta)\phi(m)
f_{\lambda}
(m,\theta){\rm d}m\,{\rm d}\theta,\] where $\tau(t-\theta)$ is the 
star formation rate (SFR)
at time $t-\theta$ in $M_{\odot}$ per time and mass units, 
$\phi(m)$ the initial mass function (IMF) defined in the interval
$[m_{\rm l},m_{\rm u}]$  and normalized to $1\ M_{\odot}$, and
$f_{\lambda}(m,\theta)$ the monochromatic flux of a star with
initial mass $m$ at wavelength $\lambda$ and at age $\theta$ since
the zero age main sequence (ZAMS) 
(null if $\theta$ exceeds the lifetime duration). A
simple discretization of both integrals leads however to oscillations of the 
emitted light (Charlot \& Bruzual 1991). 
Mainly due to the rapid evolutionary phases
such as TP-AGB or massive red supergiants, 
they can be solved with a sufficient time resolution requiring
substantial computer times (Lan\c{c}on \& Rocca-Volmerange 1996). In
fact, oscillations are observed at any time $t$ that a stellar
population of mass $m$ moves off from a stellar phase before the subsequent
population of mass $m+\delta m$ reaches the same evolutionary phase. 
Resulting oscillations present a real difficulty in simulating instantaneous bursts, 
while they are artificially smoothed with continuous star formation laws.
To avoid that problem, one possibility is to discretize only one of the two
integrals. For example (isomass method, discretization of the integral 
on mass):
\begin{eqnarray*}
F_{\lambda}(t)= & \sum_{i=1}^{p-1}\tau(t-\theta_i)\sum_{j=1}^{q-1}\phi(m_j)
(m_{j+1}-m_j) \\
& \times\int_{\theta_i}^{\theta_{i+1}}f_{\lambda}(m_j,\theta){\rm d}\theta 
\end{eqnarray*}
where $\theta_1=0$, $\theta_{p}=t$, $m_1=m_{\rm l}$, $m_{\rm q}=m_{\rm u}$
and $m_{j+1}-m_j$ is sufficiently small that equivalent phases of consecutive 
masses overlap.
An alternative is to discretize the other integral on time (isochrone method):
\[F_{\lambda}(t)=\sum_{i=1}^{p-1}\tau(t-\theta_i)(\theta_{i+1}-\theta_i)
\int_{m_{\rm l}}^{m_{\rm u}}\phi(m)f_{\lambda}(m,\theta_i){\rm d}m\]
with $\theta_{i+1}-\theta_i$ short enough, so that con\-se\-cu\-tive 
iso\-chrones 
$\int_{m_{\rm l}}^{m_{\rm u}}\phi(m)f_{\lambda}(m,\theta_i){\rm d}m$ have 
evolved little.
Both algorithms have been checked by us to give identical results.
In the following, we prefer, unlike in our previous models, the 
isochrone method partly because isochrones are directly
comparable to color-magnitude diagrams of star clusters and 
also for computational reasons.
\subsection{Stellar spectra and calibrations}
\subsubsection{The stellar library}
Although the libraries of synthetic stellar spectra become more and
more reliable, the physics of the cold stars dominating in the NIR 
($1\ \mu{\rm m}$ to $5\ \mu{\rm m}$), notably the blanketing effects that lead to
color temperatures very different to the effective ones
(Lan\c{c}on \& Rocca-Volmerange 1992), is at
the moment not taken sufficiently into account to build synthetic
spectra of galaxies. For this reason, as in our previous models,
we prefer to adopt observational libraries when possible and synthetic
spectra otherwise.
After reduction of various photometric systems to the Glass filters, standard
optical and infrared colors were derived by Bessel \& Brett (1988) for 
stars later than B8V and G0III. We have used these colors to derive fluxes at 
mean wavelengths of the infrared filters for the stars of our UV-optical 
library and fitted cubic splines to these fluxes.
Hotter star spectra are extended in the NIR with a blackbody at $T_{\rm eff}$.
The color temperatures derived in Lan\c{c}on \& Rocca-Volmerange (1992) could
be used instead, but it should make no significant difference since, at these
temperatures the blackbody is used in the Rayleigh-Jeans wavelength range.
A stellar library observed with a better resolution in the NIR with the
FTS/CFHT is in preparation.
In the mid-infrared ($\lambda > 5\ \mu{\rm m}$), we use the analytic 
extension of Engelke (1992) for stars colder than 6000 K.
For M giants, which strongly dominate in the 
NIR, we use the spectra of Fluks et al. (1994) with the temperatures
they provide. Their good resolution in 
spectral types is essential since $V-K$ increases very rapidly with decreasing
temperature.

The library from the far-UV to the NIR respects the
effective temperature of any spectral type all along the wavelength range.
Anomalous stellar spectra and wrong identifications of spectral types in the 
published libraries may produce erroneous colors and spectra of galaxies. 
For this reason, optical spectra were selected
from the library of Gunn \& Stryker (1983) according to the following procedure:
$(B-V,U-B)$, $(B-V,V-R_{\rm c})$ and $(B-V,R_{\rm c}-I_{\rm c})$ color-color diagrams for 
all the stars of the library were plotted. Least square polynomials were 
fitted
to the points, and we only selected stars in good agreement with the fits.
Effective temperatures were derived from $B-V$ according to the
calibration from Strai\v{z}ys (1992), except for M dwarfs, the temperatures
of which were calculated from $R_{\rm c}-I_{\rm c}$ according to Bessel 
(1995). 
In the far-UV (1230-3200 \AA), stellar spectra
are extracted from the IUE
ESA/NASA librairies  (Heck et al. 1984) and, after correction for 
extinction with 
the standard law of Savage \& Mathis (1979),
connected to the visible. $E_{\rm B-V}$~ is
computed from the observed $B-V$ taken from Lanz (1986) or Wesselius
et al. (1982) and the $(B-V)_0$ corresponding to the spectral type
from Strai\v{z}ys (1992). Anomalous spectra near 2000~\AA\ (especially O
stars) due to the bump of the extinction curve have been eliminated, as
well as those in strong disagreement with the slope of the UV
continuum of Kurucz (1992) for the corresponding temperature. 
In the 
extreme-UV (220-1230 \AA), we complete our spectra with Kurucz (1992) 
models for $T_{\rm eff} < 50\,000\ {\rm K}$. 
The models of Clegg \& Middlemass (1987) are finally used at all wavelengths 
for stars hotter than 50\,000 K.
Our stellar library is available in the AAS CD-ROM or on our anonymous
account.
\subsubsection{Bolometric corrections} 
We compute the bolometric corrections from
Fluks et al. (1994) spectra for M giants (getting thus a coherent set of 
spectra, temperatures and bolometric corrections), and adopt 
those given by Bessel (1995) for 
late-M dwarfs, Vacca et al. (1996) for very hot stars or else the values
tabulated by Strai\v{z}ys (1992).
The bolometric corrections that we compute from our 
spectra, since only a negligible flux should be emitted outside our wavelength
range, are in good agreement with the above values from the literature, making us 
confident that our identification in $T_{\rm eff}$ is correct and that the 
junctions between the UV, optical and NIR domains are valid.
\subsection{Evolutionary tracks}
Stars are followed from the ZAMS
to the final phases (supernovae or white dwarfs according to their
masses), including the TP-AGB, fundamental to model NIR 
spectra, and PAGB phases. 
To check the
sensitivity of spectral synthesis to evolutionary tracks, we compare
the solar metallicity tracks of  Bressan et al. (1993) (hereafter ``Padova'')
to those of Schaller et al. (1992) extended by
Charbonnel et al. (1996) (hereafter ``Geneva''). The ``Padova'' tracks 
overshoot for masses $m\geq 1\ M_{\odot}$ and use a higher ratio of 
the overshooting distance to the pressure scale height
and down to lower masses than Geneva tracks, which include overshooting
above $1.5\ M_{\odot}$ only.
Both sets 
use the OPAL opacities (Iglesias et al. 92), similar mixing lengths,
helium contents (0.28 for Padova and 0.30 for Geneva) and mass loss
rates.
We do not consider other metallicities, since these tracks already lead to
significant discrepancies (see \ref{tracescomp}) which make the
comparison of observed and synthetic spectra uncertain. 
Both sets go up to the
beginning of the TP-AGB for intermediate and low-mass stars and have
been prolonged by TP-AGB using typical luminosities and evolutionary
timescales from Groenewegen \& de Jong (1993) for stars less massive than
$6 \ M_{\odot}$. The PAGB models of Sch\"onberner (1983) and Bl\"ocker
(1995), supported by observations of planetary nebulae (Tylenda \&
Stasi\'nska 1994), are finally connected to the tracks. 

Whatever the algorithm used, interpolation between tracks requires the
identification of the corresponding points. The interpolation algorithm adopted here 
aims to conserve the released energy along any track. For Padova models,
sets of evident equivalent points are selected on consecutive mass
tracks. Considering now such points $A_i$ and
$A_j$ of track $A$ and the corresponding ones $B_p$ and $B_q$ of track
$B$, we may build a new track $B'$ replacing track $B$ with $B_i'=B_p$
and $B_j'=B_q$. Intermediate points $B_k'$ are computed iteratively
so that $E_{B'}(k,k+1)/E_{B'}(i,j)=E_A(k,k+1)/E_A(i,j)$,  where
$E(u,v)$ is the energy emitted from $u$ to $v$. 
Equivalent points are given for Geneva tracks
except in the interval $[1.7-2]\ M_{\odot}$, for which the previous procedure
has been used. For low mass stars, we use the tracks of Vandenbergh
et al. (1983). 
Except when otherwise stated, we use Padova tracks with their complements 
to the latest phases and to low mass stars, because of their higher resolution 
in mass and time and because Geneva tracks may not be interpolated after 
16 Gyr, since post-helium flash evolution is not available in the $0.8 \ M_{\odot}$ 
track.
\subsection{Nebular emission}
Gaseous nebulae are assumed to be optically thick in Lyman lines, according 
to case B recombination, the most likely 
for isolated nebulae (Osterbrock 1989).
As in our previous models, the ratio of line intensities in the
hydrogen recombination spectrum is computed for a given set of
electronic temperature and density ($T_{\rm e}=10\,000\ {\rm K}$, 
$n_{\rm e}=1\,{\rm cm^{-3}}$) of astrophysical
interest. In the NIR, the Paschen and Brackett lines were
computed, relative to Balmer lines, by Pengelly (1963) and
Giles (1977).  Other emission lines such as
${\rm H_2}$ ($2.12\ \mu{\rm m}$), He\,{\sc i} ($2.06\ \mu{\rm m}$) and [Fe{\sc ii}] ($1.6\ \mu{\rm m}$),
detected in NIR spectra of galaxies, were added in a ratio
observed in typical starbursts (Lan\c{c}on \& Rocca-Volmerange
1996). Main lines of starbursts were also added, 
following Spinoglio \& Malkan (1992). 

The nebular continuum emission coefficients in the infrared are taken
from Ferland (1980) for H\,{\sc i} and He\,{\sc ii}. H\,{\sc i} coefficients may be used
instead of He\,{\sc i} in the NIR (Ferland 1995). 
Two-photon emission coefficients are taken from Brown \& Mathews (1970) but
are negligible in the NIR. 

The number of ionizing photons is a fraction $f$ of 
the number of Lyman continuum photons
computed from our spectral library, while the rest is assumed to be 
absorbed by dust. We take $f=0.7$, 
in agreement with the values obtained by DeGioia-Eastwood (1992) for 
H\,{\sc ii} regions in the LMC.
\subsection{Extinction}
The extinction by dust which affects 
the SED of galaxies depends on 
the spatial distribution of dust and stars and on its composition,
narrowly related to the metallicity $Z$ of the ISM.
The optical depth $\tau_{\lambda}$ is related as in GRV to the column density 
of hydrogen $N_{\rm H}$ and the metallicity.

The me\-tal\-li\-ci\-ty evo\-lu\-tion $Z(t)$ is computed from Woos\-ley \& 
Wea\-ver (1995) 
SNII models without the instantaneous recycling approximation. Since the
extinction is described as a function of the global metallicity, we
neglect the yields of SNIa and intermediate and low mass stars.
Models~A are used for initial masses
$m\leq 25 \ M_{\odot}$ and 
B for $m\geq 30\ M_{\odot}$ following Timmes et al. (1995).
From ($Z=0.001$) and ($Z=0.02$) Woosley \& Weaver (1995) 
models, we may approximate the net yield $m_{\rm Z}$ in solar masses 
as follows:
\[\begin{array}{ll}
m_{\rm Z}=0 & {\rm if}\ m/M_{\odot}\leq 10.2\\ 
m_{\rm Z}=0.223m-2.27 & {\rm if}\ 10.2 \leq m/M_{\odot}\leq 18.8 \\
m_{\rm Z}=0.401m-5.62 & {\rm if}\ 18.8\leq m/M_{\odot}\leq 30.9 \\
m_{\rm Z}=8.96\,10^{-2}m+4 & {\rm if}\ 30.9\leq m/M_{\odot}.
\end{array}\]

Dust effects may be approximated in the simplest cases of 
the phase function, respectively isotropy and forward-scattering. 
Calzetti et al. (1994) proposed to model scattering effects,
with a combination of isotropic and forward-only scattering
accounting for anisotropy, for mixed dust and sources by replacing
$\tau_{\lambda}$ by the following effective depth
\[\tau_{\rm eff}(\lambda)=(h_{\lambda}(1-\omega_{\lambda})^{1/2}+
(1-h_{\lambda})(1-\omega_{\lambda}))\tau_{\lambda},\]
where the albedo $\omega_{\lambda}$ is from Natta \& Panagia (1984),
and the weight parameter $h_{\lambda}$ 
is derived by the authors from a Henyey-Greenstein phase function. 
We adopt their expression instead of the one of GRV, which corresponds 
to the case $h_{\lambda}=1$.

The geometry for the disk extinction is modelled by a uniform
plane-parallel slab as in GRV. The resulting face-on optical depth
for Sa-Sc spirals at 13 Gyr is about 0.55 in the $B$-band.
Assuming the same geometry, Wang \& Heckman (1996) deduced the face-on optical 
depth of a sample of disks from far-UV to far-infrared ratio and found
$\tau_{\rm B}=\tau_{\rm B}^{\ast}(L_{\rm B}/L_{\rm B}^{\ast})^{0.5\pm0.2}$, 
where $L_{\rm B}^{\ast}$ is 
the classical Schechter parameter of luminosity functions 
and  $\tau_{\rm B}^{\ast}=0.8\pm0.3$ the corresponding depth, 
in good agreement with our values.

To model the extinction in spheroids, we must specify the 
distribution of stars and dust. Since it is more appropriate to describe the 
inner regions of ellipticals, which are the more affected by extinction, we 
prefer to use a King model for stars rather than a de Vaucouleurs profile. 
The distribution of dust in spheroids is poorly known. Fr\"ohlich (1982)
proposed to 
describe the density of dust as a power of the density of stars:
$\rho_{\rm dust}\propto \rho_{\rm stars}^n$, and found $n\sim1/2$ 
for two ellipticals of the Coma cluster.
Witt et al. (1992) and Wise \& Silva (1996) suggest $n\sim 1/3$. 
Values $n\geq 1$ would lead
to strong color gradients in ellipticals that are not observed (Silva \& Wise 
1996).
Tsai \& Mathews (1995) obtain from the X-ray profile of three ellipticals that 
the distribution of gas is proportional to the square root of the starlight 
profile.
Assuming a constant dust to gas ratio in these galaxies, we find once again 
$n=1/2$.
We keep this value in the following and estimate the amount of extinction 
for the parameters of the model~(b) of Tsai \& Mathews (1995) which corresponds 
to a $L_{\rm B}^{\ast}$ galaxy.
We finally suppose that the galaxy geometry has not changed with time.
Introducing the core radius $r_{\rm c}$ and the outer radius $r_{\rm t}$, the
light density at a distance $r$ from the center is 
\[\rho_{\rm L}(r)=\rho_0(1+(r/r_{\rm c})^2)^{-3/2}\]
if $r<r_{\rm t}$ and 0 otherwise.
The ratio of the dust dimmed global flux $F$ of the spheroidal galaxy to the direct 
flux $F_0$ is
\[F/F_0=\frac{\int_0^{r_{\rm t}}2\pi R(\int_{-z_{\rm t}}^{z_{\rm t}}\rho_{\rm L}(r)
\exp(-\tau(R,z)){\rm d}z) {\rm d}R}
{\int_0^{r_{\rm t}}2\pi R(\int_{-z_{\rm t}}^{z_{\rm t}}\rho_{\rm L}(r){\rm d}z) {\rm d}R}\]
where $z$ and $R$ are the cylindrical coordinates, 
$z_{\rm t}=(r_{\rm t}^2-R^2)^{1/2}$, $r=(R^2+z^2)^{1/2}$ and 
$\tau(R,z)=\int_z^{z_{\rm t}}k(1+(R^2+\zeta^2)/r_{\rm c}^2)^{-3n/2}
{\rm d}\zeta$. $k$ is computed from the central optical depth 
$\tau_{\rm c}=\tau(0,-r_{\rm t})$ derived from the central column density of 
hydrogen $N_{\rm H}(0)$.
At any radius, we have
\[N_{\rm H}(R)=K\int_{-z_{\rm t}}^{z_{\rm t}}(1+(R^2+z^2)/r_{\rm c}^2)^{-3n/2}{\rm d}z,\]
where $K$ is derived from the total mass of hydrogen $M_{\rm H}$ of the galaxy.
For a helium mass fraction of 28 \% in the interstellar medium, we have 
\[M_{\rm H}=g(t)M_{\rm T}/1.4=\int_0^{r_{\rm t}}2\pi Rm_{\rm H}N_{\rm H}(R){\rm d}R,\] 
where $g(t)$ is the gas fraction, $M_{\rm T}$ the initial mass of gas of 
the galaxy and $m_{\rm H}$ the mass of the hydrogen atom.
We finally obtain for the parameters of the model~(b) of Tsai \& Mathews (1995)
\[N_{\rm H}(0)=4.7\,10^{23}g(t)\ {\rm atoms.cm^{-2}}.\]

Although the central optical depth may be very high in the early phases of
evolution of spheroidal galaxies, the extinction of the overall galaxy is only 
about $0.4$ magnitudes in the $B$-band at maximum and is negligible 
nowadays. Since neither the geometry of
stars and dust, nor the quantities and properties of dust in past E/S0 are 
known, the previous calculations should be 
considered simply as a reasonable attempt to give an order of magnitude of the 
effect of extinction in these galaxies.
\section{The star formation history of galaxies}
\label{histoire}
Scenarios of galaxy evolution aim to reproduce the spectral energy distribution
of each galaxy type on the most extended wavelength range. 
The first step is to compute the evolving SED of an instantaneous burst 
of star formation for a given IMF. 
The evolution of a real galaxy may then be described by the convolution of 
an SFR, related for example to the gas content, and of
instantaneous bursts of various ages.
\subsection{Instantaneous burst}
\subsubsection{Weight of the evolutionary tracks}
\label{tracescomp}
As shown by Charlot et al. (1996) from the comparison of the models 
of Bertelli et al. (1994), Worthey (1994) and Bruzual \& Charlot (1996),
discrepancies in the evolutionary tracks are the main sources of uncertainty 
in spectral synthesis.
Instantaneous bursts are particularly suitable to test the weight of different 
evolutionary tracks on spectral synthesis, since spectra 
\begin{figure}
\psfig{figure=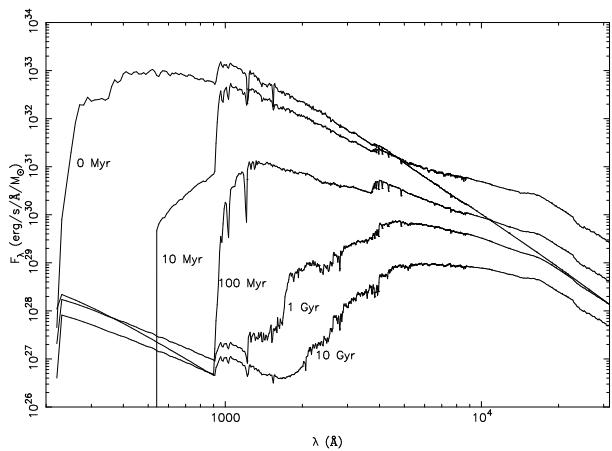,height=6.4cm,width=8.8cm,angle=-90}
\caption[]{Spectral evolution of an instantaneous burst of $1\ M_{\odot}$.
Nebular emission and extinction are not considered here.}
\label{specsursaut}
\end{figure}
and colors are not smoothed by convolution with the SFR.
In what follows, we use the polynomial form of the 
IMF obtained by Rana \& Basu (1992),
which accounts for the multiplicity of stars in the solar neighborhood.
The slope of massive stars ($m\geq 6 \ M_{\odot}$) $x\sim-1.7$ 
corresponds to the value adopted in our previous models. 
\begin{figure*}
\psfig{figure=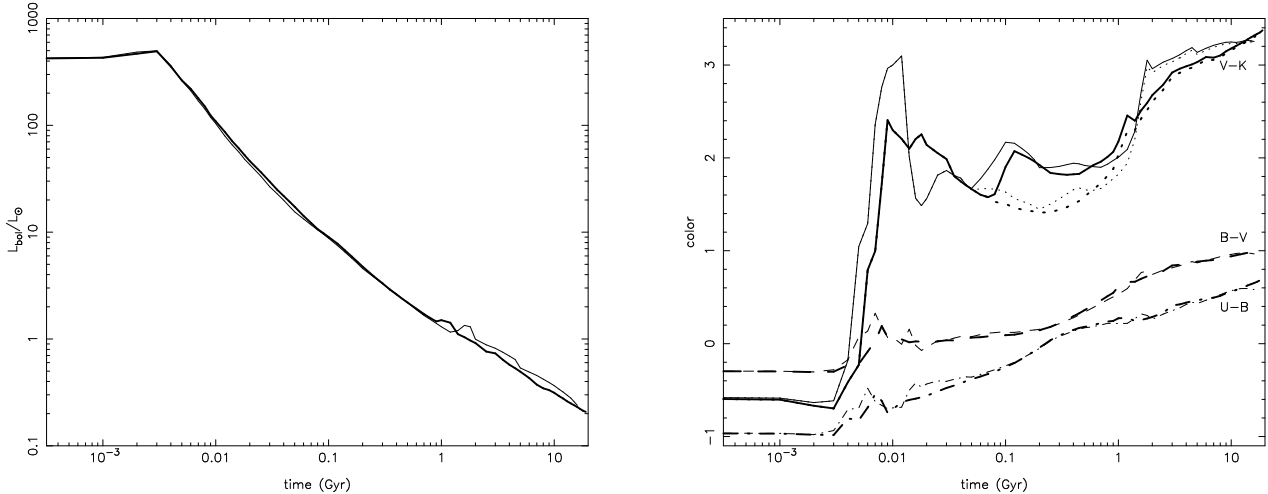,width=18cm,angle=-90}
\caption[]{Evolution of the bolometric luminosity (left) and
colors (right) of an instantaneous burst. 
Thick lines are for Padova tracks and thin ones for Geneva.
Solid: $V-K$, dashed:
$B-V$, dash-dot-dash-dot: $U-B$. The weight of thermal pulses is observed on the V-K 
curves computed without TP-AGB (dots).}
\label{tracecoulreduit}
\end{figure*}
Energy distributions of an instantaneous starburst from 200 \AA\ to
$3\ \mu{\rm m}$ are plotted at various ages in Fig.~\ref{specsursaut}.  

\begin{figure}
\psfig{figure=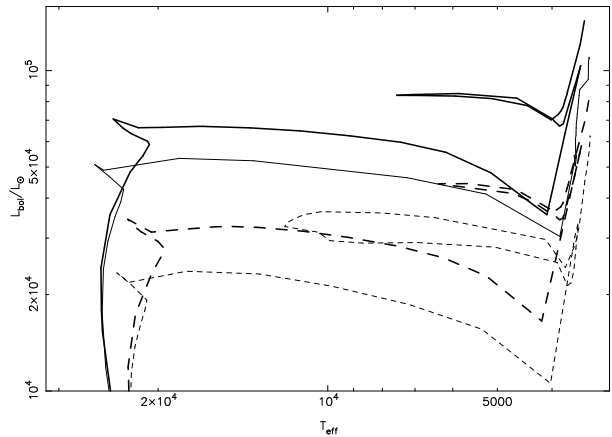,height=6.4cm,width=8.8cm,angle=-90}
\caption[]{Isochrones at 12 Myr (solid) and 18 Myr (dashed) of Padova tracks 
(thick) and Geneva tracks (thin).}
\label{isoch}
\end{figure}

We show on Fig.~\ref{tracecoulreduit} the evolution of the
bolometric luminosity with time for the tracks of Padova and
Geneva. The evolution is very similar for both sets up to 1 Gyr, where the low mass
stars of the Padova set ($m\leq 2 \ M_{\odot}$) undergo the helium flash and
provoke a bump in the bolometric luminosity. This bump is due to a flattening
of the stellar lifetime slope as a function of the
initial mass of the stars in the interval $[2-2.2]\ M_{\odot}$. 
A similar feature is obtained at 1.8 Gyr,
when the $1.7 \ M_{\odot}$ of the Geneva set also attains the helium flash. The
luminosity of the burst computed with Geneva tracks is 
systematically about $15\%$ higher after 2 Gyr than that of Padova, maybe
because of overshooting. More striking
differences may be observed on colors (see Fig.~\ref{tracecoulreduit}). 
Colors of the instantaneous starburst are derived from convolution of the 
energy distributions through filter passbands and calibration on the Kurucz 
(1992) model of Vega. Massive stars of Geneva leave the ZAMS
earlier than those of Padova but, whereas the subsequent evolution 
of $U-B$ and $B-V$ colors is 
similar, presumably because of the adjustment of the stellar evolutionary parameters on 
optical 
color-magnitude diagrams of star clusters, $V-K$ differs by as much as 0.9 
magnitude at 12 Myr and 0.75 at 18 Myr. Different treatments of the red 
supergiant phase 
are clearly responsible for this, as shown by the isochrones at 12 
and 18 Myr (see Fig.~\ref{isoch}).
While the 12 Myr isochrone of Geneva does not show any blue loop after 
the crossing of the HR diagram and is therefore redder
than that of Padova, the 18 Myr isochrone extends to much higher 
temperatures and gives bluer colors.
The color $V-K$ then evolves very slowly after 30 Myr, and both sets of tracks 
are in good agreement up to the helium flash where the bump observed in the 
bolometric 
luminosity is also visible. Geneva $V-K$ then exceeds the value of Padova by 
$\sim0.2$ mag up to 13 Gyr, where curves cross one another. The $U-B$, $B-V$ and 
$V-K$ colors of Geneva then show a flattening trend or a blueing,
the reason of which is unclear.

Although the evolution of TP-AGB is poorly known, we show on 
Fig.~\ref{tracecoulreduit} that the high-mass 
($2-7\ M_{\odot}$) TP-AGB stars may not be neglected and strongly redden the $V-K$ by 
as much as 0.6 
mag at 100 Myr. The weight of TP-AGB becomes negligible after 2 Gyr and is 
insignificant at all ages for optical colors.
Color-magnitude diagrams of optical-NIR colors should help to reduce the 
discrepancies between the tracks and to reduce the uncertainties on the 
evolution
of red supergiants, but also on the phases following the helium flash.
\subsubsection{Weight of the nebular emission}
\begin{figure}
\psfig{figure=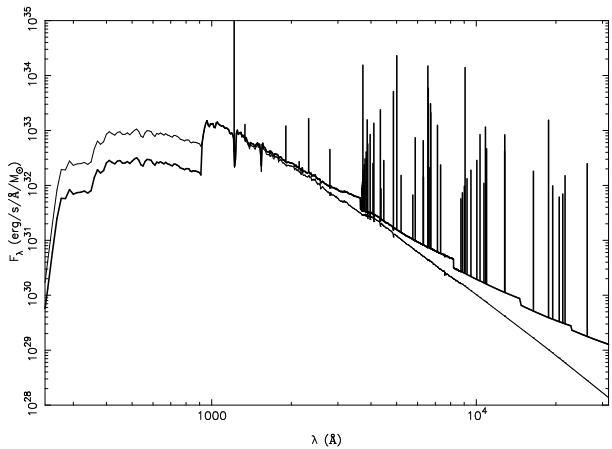,height=6.4cm,width=8.8cm,angle=-90}
\caption[]{Comparison of the spectra of an instantaneous burst at initial time 
with nebular emission (thick) and without (thin). 70\% of Lyman continuum 
photons ionize the gas and lack shortward of 912 \AA\ in the spectrum 
with nebular emission.}
\label{sursautneb}
\end{figure}

As shown in Fig.~\ref{sursautneb}, 
the inclusion of nebular emission leads to prominent features,
emission lines and discontinuities of the nebular continuum, which may not be
neglected in the
spectrum of an early instantaneous burst. To quantify the
relative effect of the nebular component, in particular in the
NIR, Table~\ref{emissneb} gives the initial fraction of burst light
through various filters, 
due to stars, lines and nebular continuum, respectively.
\begin{table}
\begin{tabular}{|c|c|c|c|}
\hline
$\lambda$ & stars & nebular continuum & emission lines\\
\hline
$U$ & 39.6 & 56.2 & 4.2\\
$B$ & 47.9 & 34.1 & 18.0\\
$V$ & 32.4 & 63.7 & 3.9\\
$R_{\rm c}$ & 20.9 & 44.5 & 34.6\\
$I_{\rm c}$ & 43.0 & 56.9 & 0.1\\
$J$ & 24.2 & 75.7 & 0.1\\
$H$ & 29.1 & 70.8 & 0.1\\
$K$ & 16.0 & 83.9 & 0.1\\
\hline
\end{tabular}
\caption[]{Weight (in \%) of stellar radiation, nebular continuum and emission 
lines at various wavelengths at initial time for an instantaneous burst.}
\label{emissneb}
\end{table}
\begin{figure}
\psfig{figure=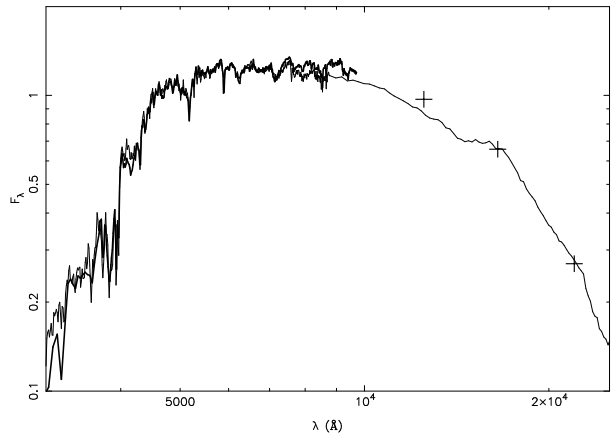,height=6.4cm,width=8.8cm,angle=-90}
\caption[]{Comparison of a 17-Gyr-old instantaneous burst (thin) with the 
E1 spectrum built by Arimoto (1996) 
(thick line and crosses) from various sources.}
\label{Arimoto}
\end{figure}
\subsubsection{The inner part of elliptical galaxies}
The E1 template, characterizing the metal-rich inner part of ellipticals, 
compiled by Arimoto (1996) from the data of
Bica (1988), Burstein et al. (1988) and Persson et al. (1979), is well 
fitted on Fig.~\ref{Arimoto}
by a 17-Gyr-old burst model from the $U$ to the $K$.
Though slightly deficient, the predicted flux in the $J$-band at solar 
metallicity is 
notably better than
in previous models of ellipticals, as noted by Arimoto (1996). The reason is 
likely the careful calibration of stellar spectra and colors 
between the $J$ and $K$ bands. 
The E1 template shows a steep UV upturn in the far UV, 
due presumably to old metal-rich stars of which discussion is beyond the 
scope of this paper. 
Its metallicity is therefore very likely
higher than solar as in the central part of typical giant ellipticals 
(Munn 1992).
Because of the age-metallicity degeneracy in the optical-NIR (Worthey 1994), 
the 
best-fitting age of 17 Gyr 
obtained with the solar metallicity model is thus only an 
upper limit. Integrated spectra on the whole galaxy of
spheroidal galaxies are however bluer in the optical-NIR and have a mean lower metallicity 
than the core (Munn 1992), allowing us to fit them at the younger age of 13 Gyr 
(see~\ref{Hubble}).

The general agreement of the E1 template with our burst spectrum
makes us confident for using our model to build various scenarios of evolution
and a new atlas of synthetic spectra (see Rocca-Volmerange \& Fioc (1996), in 
Leitherer et al. (1996b)). Timescales and resolution allow to simulate 
starbursts 
as well as evolved galaxies.
\subsection{Galaxies of the Hubble sequence}
Normal galaxies are the sum of stellar populations of different ages, well simulated with 
our basic instantaneous starburst.
Because of the insufficient knowledge of star formation physics, we prefer 
to follow the classical hypothesis of an SFR law $\tau (t)$ depending on 
the gas fraction $g$ by 
$\tau (t)=\nu g(t)$ and to explore the values of the astration rate $\nu$ 
leading to best fits of Hubble sequence galaxies.
\subsubsection{Optical-NIR colors}
\label{Hubble}
Spectral templates observed through large apertures for each Hubble type
are needed for high-redshift predictions. The litterature has
essentially published spectra obtained
through small apertures (Kinney et al. 1996) or limited to the optical wavelength range 
(Kennicutt 1992). 
Thus, we are limited to use statistical samples of colors. A coherent set
of optical-NIR colors of nearby galaxies related to morphological type 
have been obtained by Aaronson (1978) (including Huchra (1977) data) 
for an aperture $\log_{10}(A/D_0)=-0.2$. 
\begin{figure*}
\psfig{figure=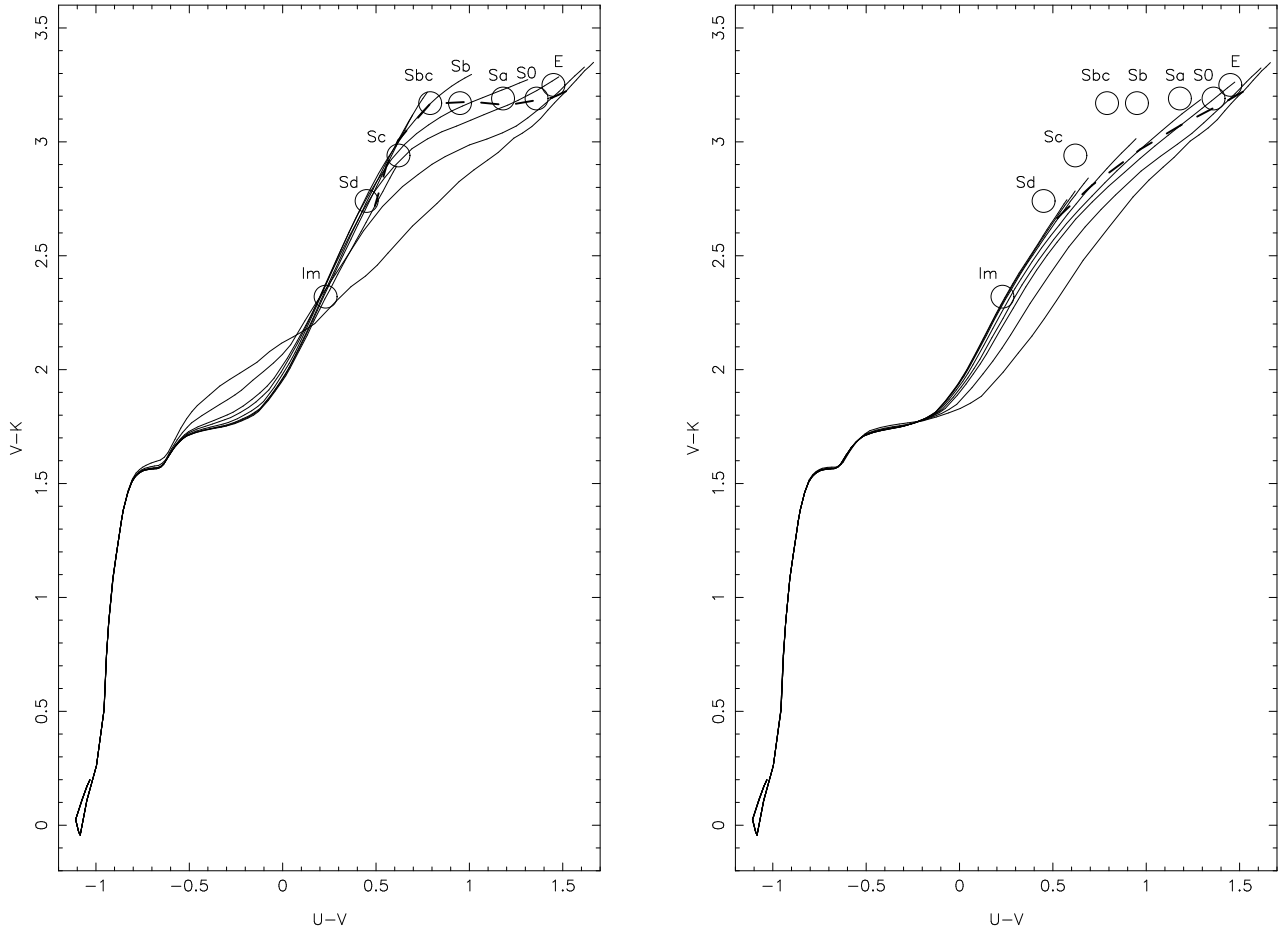,width=18cm,angle=-90}
\caption[]{Left: color evolution (solid) of the synthetic spectra for 
various star formation histories in the $(U-V,V-K)$ diagram. 
The dashed line corresponds to 13 Gyr-old
galaxies.
Extinction and nebular emission are considered.
Right: the same without extinction.
Circles are the data as in Aaronson (1978).}
\label{Aaronson}
\end{figure*}
We plot on Fig.~\ref{Aaronson} the time evolution of $U-V$ and $V-K$ colors 
for scenarios fitting the observed colors. An inclination of 1 rad,
which is the mean inclination integrated over solid angles, is
assumed for spiral galaxies. Ages of about $13\pm 2$ Gyr are obtained for 
giant spirals (Sa-Sc). Neglecting the effect of extinction would lead to
excessive ages, as shown by Fig.~\ref{Aaronson}.
The $(U-V,V-K)$ diagram shows a strong degeneracy for late
spirals. The confusion of curves corresponding to slowly-evolving SFR in this
region allows a large interval of ages for
these galaxies. Maximal age solutions are obtained for 10 Gyr-old Sd and 
4-5 Gyr-old Im. 
If we assume an increasing SFR, very similar colors 
may however be obtained at 10 Gyr for Im. 

However, for early-type galaxies, a gas-dependent SFR in a closed-box model 
with constant IMF would lead
to an excessive residual current star formation and too blue colors, 
whatever the star formation law may be. Indeed, if the bulk of the 
stars formed very early, the mass ejection of old stars $r_{\rm old}(t)$ 
will depend mainly on age and not on the detailed star formation history.
In the hypothesis of some residual star formation, 
high-mass stars which have just formed and died in releasing large amounts
of gas will also contribute to the gas content. 
This component is proportional to the current SFR and may be written
$\tau(t)r_{\rm young}$. The classical equation of evolution of the gas 
fraction in a closed model leads with previous approximations, 
for a decreasing gas fraction, to
\[\frac{{\rm d}g}{{\rm d}t}\simeq-\tau(t)+r_{\rm old}(t)+\tau(t)r_{\rm young}\le0\]
and we finally get
\[\tau(t)\ge\frac{r_{\rm old}(t)}{1-r_{\rm young}}.\]
The resulting minimal SFR gives excessively blue $U-V$ (1.35 at 16 Gyr) and,
even more importantly, UV colors. 
A possible star formation must thus be quenched to recover the colors
of normal spheroidal galaxies. However, the gas ejected by old low-mass stars 
alone since the age of 1 Gyr amounts to nearly 10\% of the mass, 
depending slightly on the IMF, anyway much more than observed (Faber \& 
Gallagher 1976). The fate of this gas is unclear and a dynamical model would be 
required to determine it, which is beyond the scope of this model.
For this reason, we simply assume the following star formation law: 
$\tau(t)=\eta \nu 
g(t)$, where $\eta=g(t)/(g(t)+g_{\rm c})$ and $g_{\rm c}$ is a threshold that
we conveniently take equal to $0.01$; $\eta$ is nearly equal to 1 at high gas 
fraction and decreases rapidly to 0 when gas rarefies, suppressing star 
formation. The remainder ($(1-\eta)\nu g$) is assumed to be expelled by late
galactic winds induced by SNIa in a low-density environment or locked in the 
formation of very low-mass stars. By relaxing the hypotheses of a closed box
or constant IMF, we obtain reasonable colors for E/S0.

\begin{figure*}
\psfig{figure=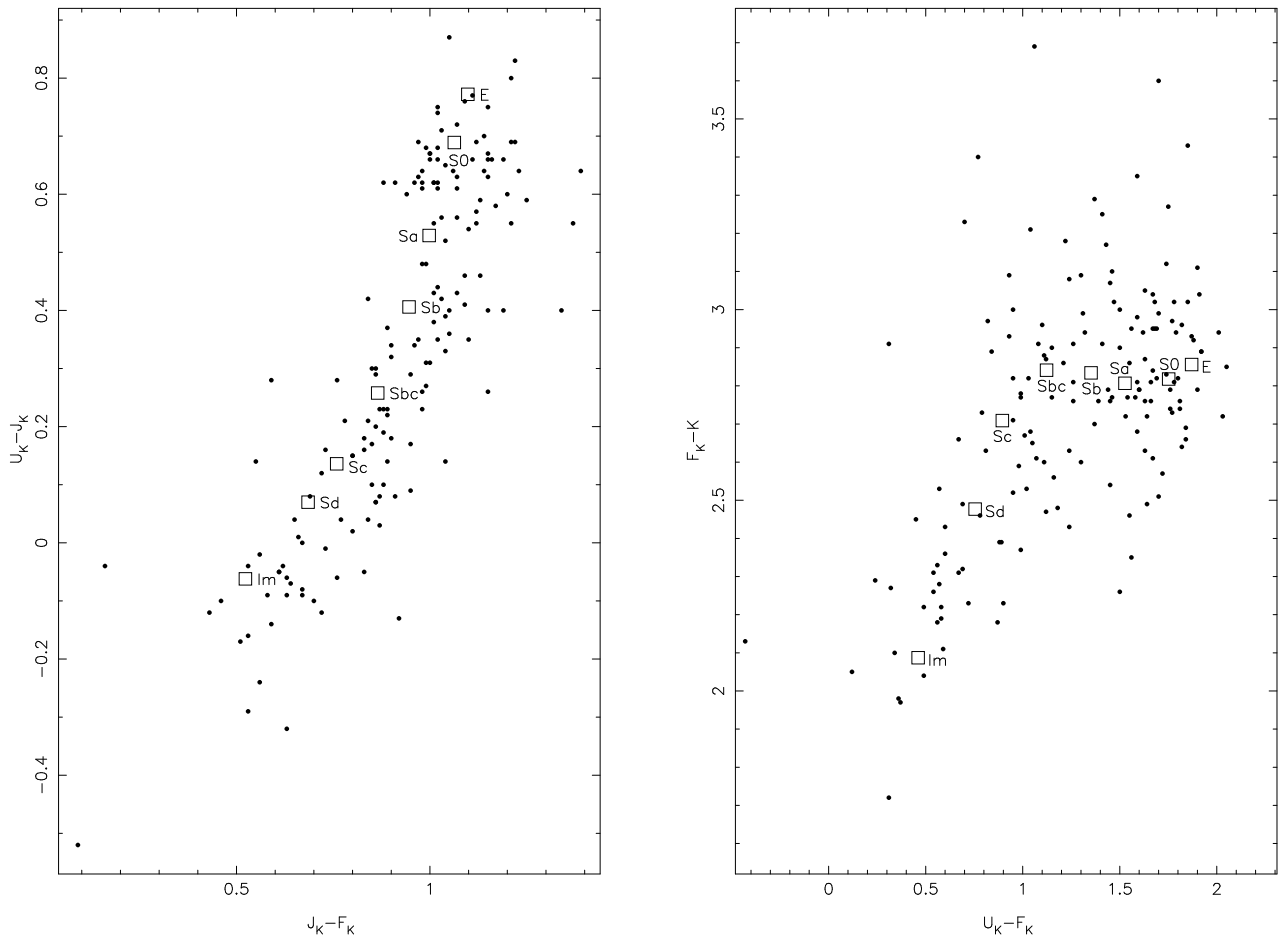,width=18cm,angle=-90}
\caption[]{Comparison to Bershady (1995) data (dots). $U$, $J$ and $F$ are
the photographic bands of Koo (1980) and Kron (1986).
Left:
$(U-J,J-F)$ diagram.
Right:
$(U-F,F-K)$ diagram.
Squares are our synthetic templates at $z=0$.}
\label{Bershady}
\end{figure*}

Our sequence of galaxy colors also compares favorably with the Bershady (1995) 
data (see Fig.~\ref{Bershady}). Although the dispersion is due partly to 
photometric uncertainties, extinction effects (inclination) and 
irregular star formation histories are certainly also important.
The slightly bluer $U-J$ of the Bershady (1995) data relative to our 
synthetic galaxies may be due to the fact that Bershady (1995)
magnitudes are computed over the whole
galaxy, whereas those of Aaronson (1978) are given for an aperture of 
$\log_{10}(A/D_0)=-0.2$.
We prefer however to keep the Aaronson (1978) values, because Bershady (1995) does 
not give a correspondence with morphological type that we need for galaxy counts.

The $B-V$ color may finally be compared with the mean values of RC3
published by Buta et al. (1994) to check that the identifications of
the morphological types of Aaronson (1978) 
are correct. Since these are face-on corrected values, we
correct spiral RC3 colors to an inclination of 1 rad corresponding to a value 
of $\log_{10}R_{25}\sim 0.25$ and gather RC3 types to agree with our types 
when necessary. Differences never exceed 0.05 mag.

The characteristics and the main colors of the computed spectra as well as the 
RC3 values are given in Table~\ref{coulmod}.
\begin{table*}
\begin{tabular}{|c|c|c|c|c|c|c|c|c|c|c|c|c|c|c|c|}
\hline
Model type    & E     & S0       & Sa   & Sb   & Sbc  & Sc    & Sd    & Im     \\
\hline
RC3 types      & -5,-4 & -3,-2,-1 & 0,1  & 2,3  & 4    & 5     & 6,7,8 & 9,10\\
$\nu$ ($10^{-3}\ M_{\odot}.{\rm Myr}^{-1}$) & 2 & 1 & 0.5 & 0.35 & 0.2 & 0.1 & 0.07 & 0.05\\
age (Gyr)    & 13    & 13       & 13   & 13   & 13   & 12    & 10    & 4.5\\
$(B-V)_{\rm RC3}$ & 0.91  & 0.90     & 0.83 & 0.75 & 0.65 & 0.59  & 0.50  & 0.44\\
$B-V$         & 0.95  & 0.92     & 0.84 & 0.78 & 0.69 & 0.59  & 0.53  & 0.39\\
$U-B$         & 0.57  & 0.51    & 0.34 & 0.23 & 0.09 & 0.00  & -0.06 & -0.16\\
$V-K$         & 3.23  & 3.19     & 3.16 & 3.18 & 3.17 & 2.96  & 2.75  & 2.31\\
$V-R_{\rm c}$        & 0.61  & 0.59     & 0.57 & 0.56 & 0.54 & 0.49  & 0.45  & 0.36\\
$V-I_{\rm c}$        & 1.26  & 1.24     & 1.21 & 1.20 & 1.16 & 1.06  & 0.97  & 0.78\\
$J-H$         & 0.75  & 0.75     & 0.75 & 0.76 & 0.77 & 0.75  & 0.71  & 0.64\\
$H-K$         & 0.18  & 0.18     & 0.18 & 0.19 & 0.20 & 0.19  & 0.17  & 0.15\\
$M/L_{\rm B}$ ($M_{\odot}/L_{\rm B_{\odot}})$ & 6.92 & 6.15 & 5.02 & 4.60 & 
4.25 & 3.87 & 3.74 & 3.67\\
\hline
\end{tabular}
\caption[]{Evolutionary parameters and colors of the synthetic templates.}
\label{coulmod}
\end{table*}
\subsubsection{Optical spectra}
\begin{figure*}
\psfig{figure=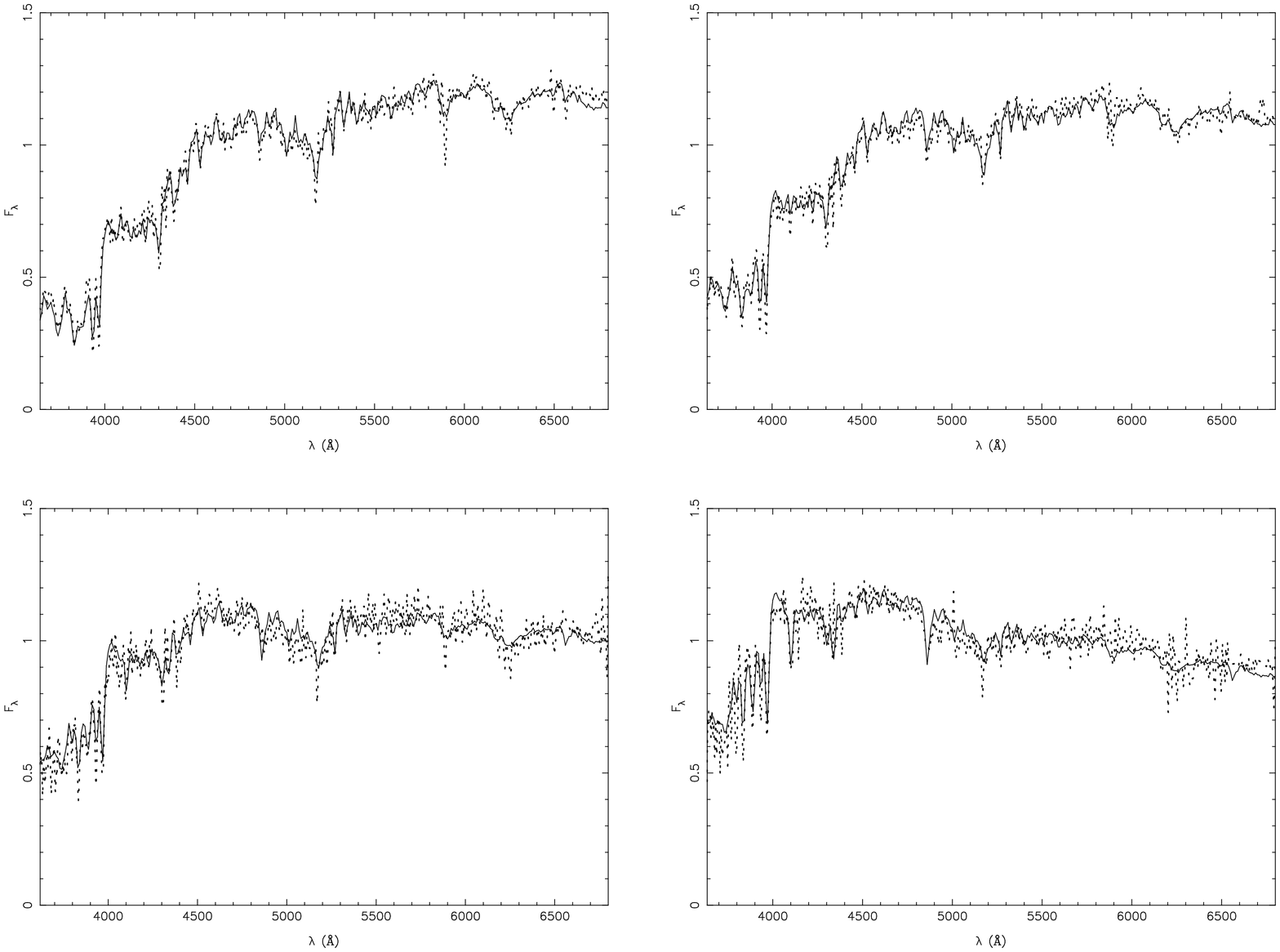,width=18cm,angle=-90}
\caption[]{Comparison of our synthetic spectra at $z=0$ (solid) to Kennicutt 
(1992) spectra (dotted).
Top left: synthetic elliptical vs. NGC 3379 (RC3 type=-5).
Top right: synthetic Sa vs. NGC 3368 (RC3 type=2).
Bottom left: synthetic Sbc vs. NGC 3147 (RC3 type=4).
Bottom right: synthetic Sd vs. NGC 6643 (RC3 type=5).}
\label{Kennicutt}
\end{figure*}
We may compare our spectra in the optical range with those 
of Kennicutt (1992). Our spectral continua are in very close agreement with 
that sample (Fig.~\ref{Kennicutt}), after correction for redshift and deletion of emission lines
by Galaz \& de Lapparent (1996). In particular, typical features such as the Balmer jump
and the MgI and TiO lines are surprisingly well reproduced.
\section{Calibration of bright counts}
\label{comptages}
Because they do not depend on the cosmological parameters, bright galaxy counts
($15<b_{\rm j}<19$) are a straightforward constraint on spectral evolution models
and notably on the reliability of the $z=0$ templates.  
Two related problems concerning bright counts arose in previous analyses 
(Guiderdoni \& Rocca-Volmerange 1990; Pozetti et al. 1996), namely the 
slope at bright magnitudes and the normalization of the luminosity function.
Maddox et al. (1990) have found in the APM survey a steeper slope of number 
counts between $b_{\rm j}=15$ and $b_{\rm j}=19$ than allowed by pure 
luminosity evolution models, implying a much more rapid
evolution at low redshift. As a result, when model predictions are 
calibrated on observed counts in the brightest bins, a strong deficit of 
predicted counts relative to observed ones is obtained at faint blue 
magnitudes. Although the corresponding low normalization of the luminosity 
function is in agreement with Loveday et al. (1992), most models have preferred until now
to normalize the counts at fainter magnitudes ($b_{\rm j}\sim19$), where 
evolution may become non-negligible, to recover agreement with
faint counts.

This is clearly a crucial problem for the 
reliability of evolution models.
However, recent observational results do not confirm the conclusions of 
Maddox et al. (1990). 
The re-analysis of the photometry of APM data by Metcalfe et al. (1995) 
and the counts of Bertin \& Dennefeld (1996) favor a flatter slope and 
a high normalization of the luminosity function. 
To clarify this controversy, a multispectral analysis of bright counts
is needed, highlighting by different weights of old and young 
stellar populations in the various bands, whether a strong evolution has 
happened recently or not. The NIR, in particular,
is dominated by slowly-evolving old stellar populations and should therefore 
be more sensitive to number density evolution than to recent star formation.
Moreover, in this wavelength range, the 
k-corrections for different types are very similar, contrary to the UV. 
Finally, it is less affected than the blue by the 
uncertainties in the luminosity functions of late-type galaxies. 
For these reasons, predictions in the $K$-band are more reliable than at 
shorter wavelengths.
The evolutionary spectra
built by PEGASE are particularly useful for this purpose, thanks to their 
large continuous range of wavelength and valid fits of
nearby observational templates.

\begin{figure}
\psfig{figure=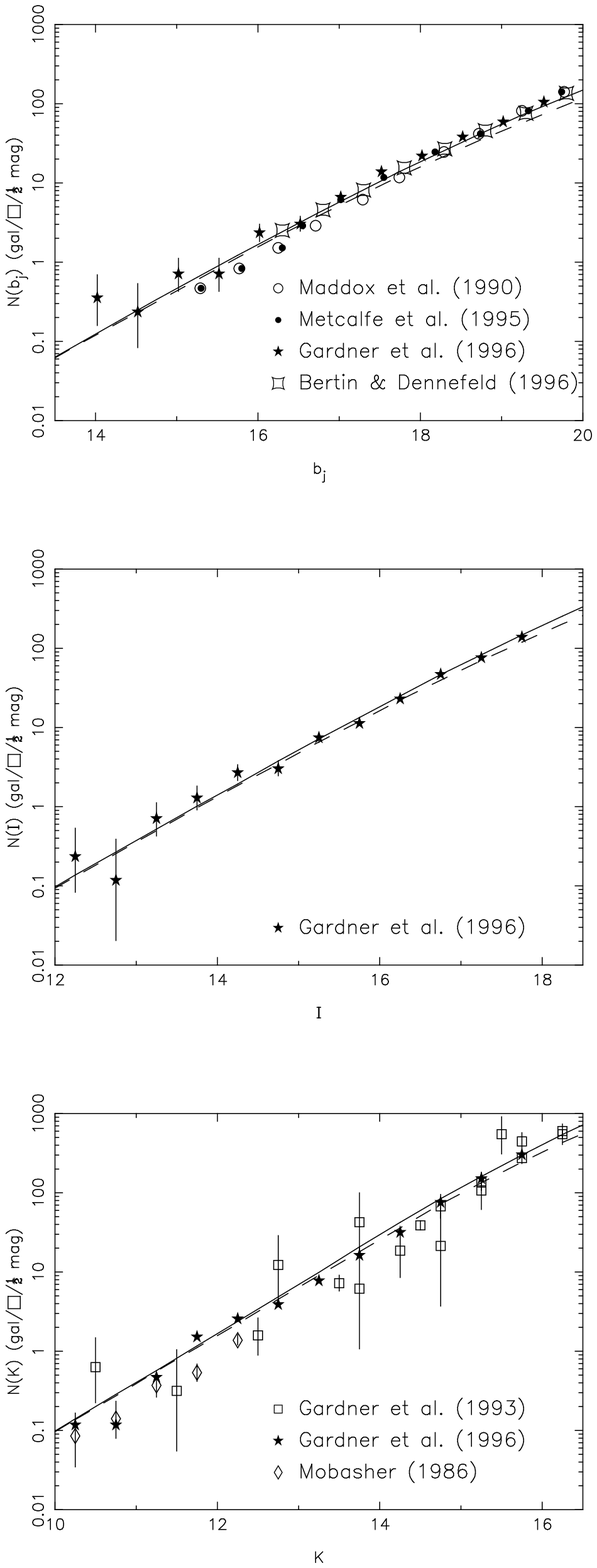,width=7.9cm}
\caption[]{Galaxy counts in $b_{\rm j}$, $I$ and $K$ with the Marzke et al. (1994) 
luminosity function. 
Cosmological parameters are $H_0=65\ {\rm km.s^{-1}.Mpc^{-3}}$, $q_0=0.05$ and 
$\Lambda_0=0$.
The solid (resp. dashed) lines are our predicted counts with (resp.
without) evolution.}
\label{traceN_BIKreduit}
\end{figure}

We have modelled galaxy counts in the 3 bands $b_{\rm j}$, $I$ and $K$ (Fig.~\ref{traceN_BIKreduit}).
The adopted luminosity function is that of Marzke et al. (1994), after correction of the Zwicky
magnitudes $M^{\ast}_{\rm Z}$ by $B^{\ast}-M^{\ast}_{\rm Z}=-0.29$, the mean
value obtained by Efstathiou et al. (1988), to convert them in $B$ magnitudes. 
Characteristics of the luminosity functions are given in Table~\ref{fonclum}. 
When necessary, we redistribute the types given for the LF in our standard 
types. We take $H_0=65\,{\rm km.s^{-1}.Mpc^{-1}}$ in agreement with Tanvir et al. (1995).
An open cosmology ($q_0=0.05$, $\Lambda_0=0$) is assumed, leading to an age of the universe 
of 13.5 Gyr greater than the age of our older standard galaxies, but does not in any way 
affect the simulation of bright counts. 
Redshifts of formation in this cosmology
(see Table~\ref{fonclum}) are taken in agreement, within the uncertainties,
with the ages of the reference model spectra.
We finally normalize our counts on Bertin \& Dennefeld (1996) counts at 
$b_{\rm j}=16$.

We obtain good agreement with Gardner et al. (1996) {\em multispectral} 
bright counts in the 3 bands $b_{\rm j}$, $I$ and $K$ for the same 
value of normalization, confirming that the evolution scenarios and colors 
of the galaxies dominating bright counts are correct. Moreover, 
the value of $\phi^{\ast}=23.8\,10^{-3}h^3\ {\rm Mpc}^{-3}$ from modeling is fully consistent with 
the normalization $(20.1\pm5.0)10^{-3} h^3 {\rm Mpc}^{-3}$ of the Marzke et al. (1994) luminosity 
function that we use.
\begin{table}
\begin{tabular}{|c|c|c|c|c|}
\hline
Type	& $z^{\rm for}_i$ & $B^{\ast}_i$	& $\alpha_i$	& 
$g_i=\phi^{\ast}_i/\sum \phi^{\ast}_i$\\
\hline
E	& 20		& -19.52	& -0.85		& 0.066\\
S0	& 20		& -19.03	& -0.94		& 0.250\\
Sa	& 5		& -19.03 & -0.94	& 0.083\\
	& 		& -19.01 & -0.58	& 0.095\\
Sb	& 5		& -19.01	& -0.58		& 0.191\\
Sbc	& 5		& -19.01	& -0.58		& 0.096\\
Sc	& 5		& -19.10	& -0.96		& 0.064\\
Sd	& 2		& -19.10	& -0.96		& 0.129\\
Im	& 0.5		& -19.08	& -1.87		& 0.026\\
\hline
\end{tabular}
\caption[]{Redshifts of formation and luminosity functions 
(Schechter parametrization) of our standard synthetic spectra. $B^{\ast}$ is  
given for $H_0=100\ {\rm km.s^{-1}.Mpc^{-1}}$. Two LFs are given for Sa  
and should be added. The reason for this is that 
the S0 type of Marzke et al. (1994) includes S0/a,
while these are included in our synthetic Sa. Part of the S0 LF is for this
reason transferred to the synthetic Sa.}
\label{fonclum}
\end{table}

Predictions at fainter magnitudes in various cosmologies 
and constraints from redshift and color 
distributions as well as correlation functions
will be extensively discussed in a future paper.
\section{Discussion and conclusion}
The understanding of galaxy evolution requires models of spectral
evolution. The model proposed here benefits from many improvements and new input data, 
in particular in the NIR. It becomes sufficiently reliable 
to be entirely published as a useful tool for many evolution studies.
The wide wavelength range (220 \AA--$5\ \mu{\rm m}$)
of the library of stellar spectra
and the capacity of the algorithm to follow the evolution on very rapid 
(1 Myr) and very long timescales (20~Gyr)
provide constraints on the weights of various stellar populations in 
starbursts and evolved galaxies. 
The inclusion of extinction, notably in E/S0, and nebular emission has been 
attempted in consistency with the evolution of stars. 
A library of standard synthetic evolving spectra has been built and shows a
fair agreement with $z=0$ galaxies from observed optical spectra and 
colors in the near-UV, optical and NIR. 
Several samples of templates are used to define a set of reference synthetic 
spectra at $z=0$, adopted as typical of 8 spectral types of the Hubble 
sequence. 
An age of about 13 Gyr is derived from optical-NIR colors for spheroidals and
early spirals and decreases for later types.

Predicted galaxy counts agree with most recent optical and NIR bright counts
as well as with new determinations of the normalization of the luminosity 
function, confirming 
that the evolution of nearby galaxies is well described by pure luminosity
evolution models.

Uncertainties in the stellar evolutionary tracks are discussed in this paper,
and we tried to reduce at maximum those related to stellar 
atmospheres. Other main uncertainties concern the effects of extinction and
metallicity. The optical thickness of spirals, as well as the state of the ISM 
and the resulting extinction in ellipticals, are still a matter of debate. 
The influence of the metallicity on extinction curves, stellar spectra 
and evolutionary tracks could be important
in metal-rich ellipticals as well as metal-poor H\,{\sc ii} galaxies.
Finally, the star formation (SFR, IMF) in galaxies should be derived from a 
chemo-dynamical model.

In spite of these uncertainties, this model should prove useful to the
astronomical community. The code sources, input data, especially 
our selected stellar library\footnote{We also propose another library of 
intermediate resolution (1.4 \AA) optical stellar spectra selected from 
Jacoby et al. (1984).}, 
and the atlas of evolving synthetic spectra of standard galaxies 
(spectra, colors, equivalent widths of emission lines, numbers of SNII)
with corresponding k- and e-corrections are available on an AAS
CD-ROM (see Fioc \& Rocca-Volmerange (1996) and  Rocca-Volmerange \& Fioc 
(1996) in Leitherer et al. (1996b)) 
and the latest version corresponding to this paper may be 
obtained at addresses given in the abstract. It may be easily used by anyone, even 
unfamiliar with spectral evolution questions, but also, 
thanks to its structure, adapted by those interested 
in this subject. We hope to allow fruitful comparisons with
other existing models.
Further improvements will also be available by ftp. 
Comments, questions and suggestions may be sent to \mbox{\em pegase@iap.fr}.

\end{document}